\documentclass[12pt]{iopart}
\usepackage{times}
\usepackage{t1enc}
\usepackage[english]{babel}
\usepackage{graphicx}
\usepackage{relsize}
\usepackage{iopams}
\usepackage{natbib}
\usepackage{mathptmx}
\usepackage{xcolor}

\eqnobysec

\begin{document}
\title{Corneal retardation time as an ocular hypertension disease indicator}

\author{Oscar del Barco$^{1,*}$, Francisco J \'{A}vila$^{2}$,
Concepci\'{o}n Marcell\'{a}n$^{2}$ and Laura Rem\'{o}n$^{2}$}

\address{$^1$ Laboratorio de \'{O}ptica, Instituto
Universitario de Investigaci\'{o}n en \'{O}ptica y
Nanof\'{i}sica, Universidad de Murcia, Campus de
Espinardo, E-30100, Murcia, Spain}
\address{$^{2}$ Departamento de F\'{i}sica Aplicada,
Universidad de Zaragoza, E-50009, Zaragoza, Spain}
\address{$^*$ Author to whom any correspondence should be addressed.}
\ead{obn@um.es}

\begin{abstract}
{\it Objective.} A detailed analysis of the corneal
retardation time $\tau$ as a highly related parameter to
the intraocular pressure (IOP), and its plausible
role as an indicator of ocular hypertension disease.
{\it Approach.} A simple theoretical
expression for $\tau$ is derived within the
corneal viscoelastic model of Kelvin-Voigt with
3 elements. This retardation time can be
easily calculated from the well-known signal
and pressure amplitudes of non-contact tonometers
like the Ocular Response Analyzer (ORA). Then, a
population-based study was performed where 100 subjects
aged from 18 to 30 were analyzed (within this group,
about 10\% had an elevated IOP with more than 21 mmHg).
{\it Main results.} A clear relationship between the
corneal retardation time and the corneal-compensated
intraocular pressure ($\textrm{IOP}_{\rm cc}$)
was found, underlying the risk for ocular hypertensive
(OHT) subjects with lower $\tau$ values to develop
hypertension illnesses (due to the inability of
poorly viscoelastic corneas to absorb IOP fluctuations,
resulting in probable optic nerve damage).
{\it Significance.} Our results might provide an useful tool
to systematically discern which OHT patients
(and even those with normal IOP values)
are more likely to suffer glaucoma progression
and, consequently, ensure an early diagnosis.
\end{abstract}


\maketitle

\section{Introduction}

Corneal biomechanics (CB) is a branch of
biophysical sciences that deals with deformation
and equilibrium of corneal tissue when any external
force is applied. In this sense, the mechanical
properties of the corneal tissue depend on the specific
organization of fibres, cells and ground substance
within the structure. Collagen in Bowman's layer and
stroma make a significant contribution to
corneal elasticity, whereas the ground substance
would give the viscous behaviour (Garcia 2014).
The increasing interest in CB is due to,
among others, its role in the detection
and management of ectatic disease
(Ortiz \emph{et al.} 2007, Gonz\'{a}lez \emph{et al.} 2008,
Roy and Dupps 2011, Ambr\'{o}sio \emph{et al.} 2017,
Padmanabhan and Elsheikh 2023) and an accurate
estimation of IOP to manage pathological
diseases such as glaucoma
(Liu and Roberts 2005, Susanna \emph{et al.} 2019,
Asejczk-Widlicka \emph{et al.} 2019,
Consejo \emph{et al.} 2019, Chan \emph{et al.} 2021,
Catania \emph{et al.} 2023).

There are different material models which
describe with more or less accuracy the corneal
biomechanics. In this respect, we can mention
the visco-hyperelastic model, where a highly
nonlinear elastic response is achieved when
very large strains are applied
(Ariza-Gracia \emph{et al.} 2015,
Whitford \emph{et al.} 2018, Liu \emph{et al.} 2020),
the viscoelastic model (i.e., the material's elastic
stress-strain relationship depends on the strain rate)
(Fraldi \emph{et al.} 2016, Maczynska \emph{et al.} 2019)
or the finite element methods, where a complete 3D
model of the cornea is designed to study
its mechanical behaviour (S\'{a}nchez \emph{et al.} 2014,
Simonini \emph{et al.} 2016).
On the other hand, one-dimensional (1D)
rheological models have been useful to describe the
viscoelastic properties of the cornea
(Glass \emph{et al.} 2008, Han \emph{et al.} 2014,
Jannesari \emph{et al.} 2018), though they are not
meant to study the 3D corneal deformation.

In this regard, 1D models consist of parallel
and/or series combinations of springs and dashpots
which mimic the elastic and/or viscous character
of the cornea. Thus, a Kelvin-Voigt model with an
additional spring can reproduce the instantaneous
deformation of the cornea (Glass \emph{et al.} 2008),
while a four-element viscoelastic model
(i.e., the Burgers model) has also been selected
for modeling the corneal biomechanics
(Jannesari \emph{et al.} 2018).
More recently, a more sophisticated rheological
model that takes into account the elastic and viscous
effects of cornea, crystalline lens and the
whole eyeball has been reported
(Jimenez-Villar \emph{et al.} 2022).
Although more complex models with a
greater number of elements (i.e., springs and/or dashpots)
should be more accurate in corneal modeling (Kok \emph{et al.} 2014,
Jannesari \emph{et al.} 2018),
these approaches might not present a unique
mathematical solution, due to the higher-order
differential equations inherent in these
models. So, as clearly stated by
Jannesari \emph{et al.} (2018),
rheological models combining simplicity
with accuracy are desired.

Furthermore, it has been amply demonstrated
that CB influences IOP measurements
(Medeiros and Weinreb 2006,
Grise-Dulac \emph{et al.} 2012,
Brown \emph{et al.} 2018).
For that matter, the corneal-compensated
intraocular pressure provided by the
non-contact tonometer ORA is less influenced by corneal
biomechanics (Medeiros and Weinreb 2006,
Hager \emph{et al.} 2008, Lee \emph{et al.} 2019),
so it might be a reliable
parameter to characterize OHT subjects.
As it is well-known, an elevated IOP is
the major risk factor for
developing glaucoma (De Moraes \emph{et al.} 2012,
Matlach \emph{et al.} 2019), however,
this is not the unique factor. It has been reported
glaucomatous damage at low IOP values (Anderson 2003),
whereas no significant glaucoma progression
has been found at IOPs greater than 22 mmHg
(Kass \emph{et al.} 2002).

In this connection, the gold standard
method widely used by
ophthalmologists to evaluate structural changes
in the optic nerve head (ONH) or the retinal nerve
fiber layer (RNFL) and assist in the diagnosis of
glaucoma has been the fundus photography
(Chakrabarti \emph{et al.} 2016). The main
advantage of this technique is its simplicity
and cost-effectiveness, despite the clinical
examination of ONH and RNFL is subjective
and qualitative, leading to considerable
intra- and interobserver variability
in assessing the ONH among qualified specialists.
Alternative methods such as optical coherence tomography
(OCT) (Geevarghese \emph{et al.} 2021), scanning laser
polarimetry (SLP) (Lemij and Reus 2008), and confocal
scanning laser ophthalmoscopy (CSLO) (Yaghoubi
\emph{et al.} 2015) have been developed
to evaluate nerve fiber loss and optic disc
changes in glaucoma. Nonetheless, these retinal
imaging instruments are costly and present some
drawbacks, among them, the susceptibility of CSLO to
interobserver variabilities or the inability
of SLP method to provide both RNFL and ONH data.
Additionally, selective perimetry techniques such
as short-wavelength automated perimetry (SWAP) and
frequency-doubling technology (FDT) perimetry have been
extensively studied as adjuncts to standard automated
perimetry evaluation (Sharma \emph{et al.} 2008).

Accordingly, the aim of this work is to
yield a reasonable indicator related to the
viscoelastic corneal quality, which might be useful
to discern which OHT subjects are more probable
to develop ocular hypertensive disorders such as glaucoma.
This parameter is the corneal retardation time
$\tau$, that is, the time in
which about 63\% of the final corneal strain
is determined (Brinson and Brinson 2008,
Jannesari \emph{et al.} 2018), and might
serve as an indicator of how elastic or viscous
a cornea should be. In other words, this metric
would measure the cornea's ability to absorb IOP
fluctuations. As we will show in this article,
the $\tau$ parameter might explain why some
OHT subjects (and even those with normal IOP values)
should undergo glaucoma progression, while others not.

The paper is organized as follows. In Section 2 we
describe our 1D corneal
viscoelastic model to derive a practical
expression for the corneal retardation time $\tau$,
as a function of the corneal applanation pressures
and their first derivatives. A detailed
explanation of our methods to calculate the $\tau$
parameter is performed in Section 3, and our
corneal retardation results concerning a population
of 100 healthy young subjects is presented
in Section 4. Finally, we discuss and
summarize our results in Section 5.

\section{Theoretical calculation of the corneal retardation time}

Let us first introduce the theoretical model for the
corneal biomechanics, in order to derive
a simple and useful expression
for the corneal retardation time $\tau$
(i.e., our crucial parameter which might be used
as a plausible OHT disease indicator).

\begin{figure}
\begin{center}
\includegraphics[width=.77\linewidth]{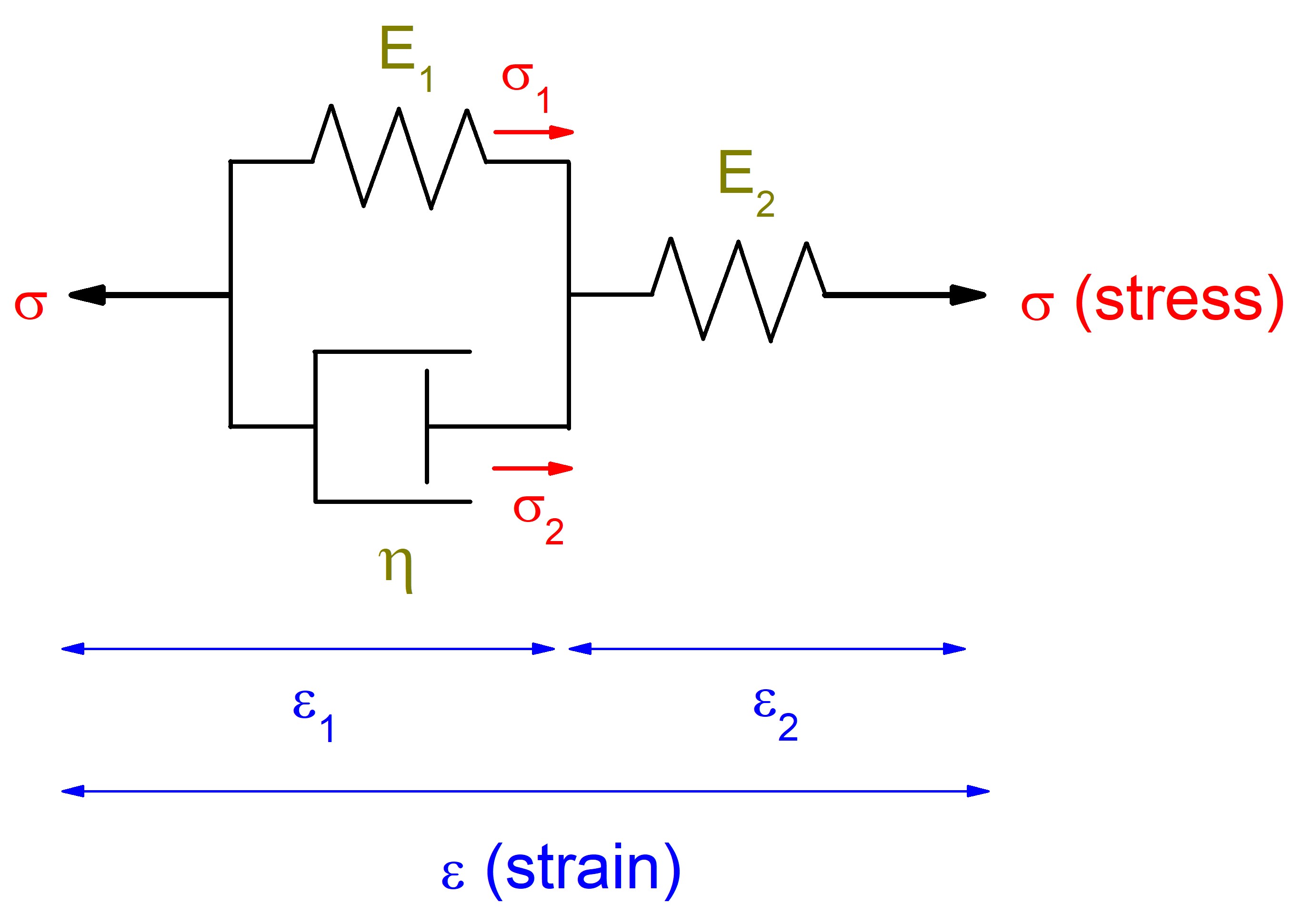}
\caption{Kelvin-Voigt viscoelastic model of three elements
for the corneal biomechanics. This configuration allows
an instantaneous deformation of the cornea through spring
$E_{2}$ and a progressive deflection though the
parallel component of dashpot $\eta$ and spring $E_{1}$.
Upon unloading, $E_{2}$ contracts immediately and
the left-hand spring slowly shortens, being held
back by the dashpot.}
\label{fig1}
\end{center}
\end{figure}

When loaded, the cornea demonstrates some
instantaneous deformation (purely elastic behavior)
followed by a progressive viscoelastic
deflection. This trend can be fairly described by the
Kelvin-Voigt viscoelastic model of three elements (KVM)
(please, see figure \ref{fig1}) where the dashpot $\eta$
symbolizes the time-dependent viscous resistance
to the applied force, and springs $E_{1}$ and $E_{2}$
mimic the purely elastic behavior
(Glass \emph{et al.} 2008). When a stress
$\sigma$ is applied, a corneal strain $\epsilon$ is induced.
This configuration allows an instantaneous deformation
of the cornea through spring $E_{2}$. More precisely,
the right-hand spring $E_{2}$ stretches
immediately upon loading. Then, the dashpot $\eta$
then takes up the stress, transferring
the load to the second spring $E_{1}$ as it
slowly varies over time. Upon unloading, $E_{2}$
contracts immediately and the left-hand spring
slowly shortens, being held back by the dashpot.

Though more sophisticated 1D
rheological models have been recently studied
(Jannesari \emph{et al.} 2018,
Jimenez-Villar \emph{et al.} 2022),
we have chosen the KVM over other viscoelastic
approaches (such as the Zener or Burgers models)
for two reasons: its ability to
mimic the corneal response to an applied force
(as above-mentioned, an instantaneous deformation
followed by a progressive viscoelastic deflection)
and the limited number of independent
variables (thus, reducing its mathematical complexity).
As currently explained by Torres \emph{et al.} (2022),
the Kelvin-Voigt model is quite appropriate and
straightforward to characterize the viscoelasticity
of the cornea, as well as it has
been experimentally validated with artificial
phantom corneas (Glass \emph{et al.} 2008).

Our KVM relates the applied stress ($\sigma$) to the
corneal strain ($\epsilon$) through the following
set of equations (Kelly 2013)
\begin{eqnarray}\label{seteqs}
& & \sigma = \sigma_{1} + \sigma_{2} \qquad \epsilon =
\epsilon_{1} + \epsilon_{2} \qquad
\sigma_{1} = E_{1} \epsilon_{1}
\nonumber\\
& & \sigma = E_{2} \epsilon_{2} \qquad
\quad \sigma_{2} = \eta \dot{\epsilon}_{1},
 \end{eqnarray}
and $\dot{\epsilon}_{1}$ corresponds to the strain rate
of the parallel elements.
Assuming the cornea to be axisymmetric, a single elastic
constant should govern corneal behavior
(Glass \emph{et al.} 2008),
so we can identify $E_{1} = E_{2} = E$.
After Laplace transforming, the constitutive
relation can be written as (Kelly 2013)
\begin{equation}\label{ODE}
E \epsilon + \eta \dot{\epsilon} =
2\sigma + \tau \dot{\sigma},
\end{equation}
where $\dot{\sigma}$ is the stress rate
and $\tau=\eta/E$ stands for the corneal
retardation time. The latter parameter
describes the time dependent response of
the cornea with respect to the applied load.

On the other hand, non-contact tonometers
such as ORA applanate the cornea in two instants,
when the strain is minimum.
Consequently, the strain rate cancels at
these applanation moments and
$\dot{\epsilon}=0$. So, from
equation (\ref{ODE}) we can write for
the first applanation time $t_{\rm ap,1}$
\begin{equation}\label{fstapp}
E \epsilon(t_{\rm ap,1})= 2\sigma(t_{\rm ap,1}) +
\tau \dot{\sigma}(t_{\rm ap,1}) =
-2 |\sigma(t_{\rm ap,1})| +
\tau |\dot{\sigma}(t_{\rm ap,1})|,
\end{equation}
where it is assumed a compressive stress
($\sigma(t_{\rm ap,1}) < 0$)
during the load stage ($\dot{\sigma}(t_{\rm ap,1})>0$).
Moreover, for the second applanation time
(also a compressive regime with
$\sigma(t_{\rm ap,2}) < 0$), we have
\begin{equation}\label{secapp}
E \epsilon(t_{\rm ap,2})= 2\sigma(t_{\rm ap,2}) +
\tau \dot{\sigma}(t_{\rm ap,2}) =
-2 |\sigma(t_{\rm ap,2})| -
\tau |\dot{\sigma}(t_{\rm ap,2})|,
\end{equation}
and now, during the unload process,
$\dot{\sigma}(t_{\rm ap,2})<0$.
Provided that the strain at applanation is the
same for both load-unload processes
(i.e., $\epsilon(t_{\rm ap,1}) =
\epsilon(t_{\rm ap,2})$), we obtain the
following expression for the corneal
retardation time $\tau$ from equations
(\ref{fstapp}) and (\ref{secapp})
\begin{equation}\label{taucor}
\tau = \frac{2 \left(|\sigma(t_{\rm ap,1})|-
|\sigma(t_{\rm ap,2})|\right)}
{|\dot{\sigma}(t_{\rm ap,1})| +
\dot{\sigma}(t_{\rm ap,2})}.
\end{equation}

\begin{figure}
\begin{center}
\includegraphics[width=.45\linewidth]{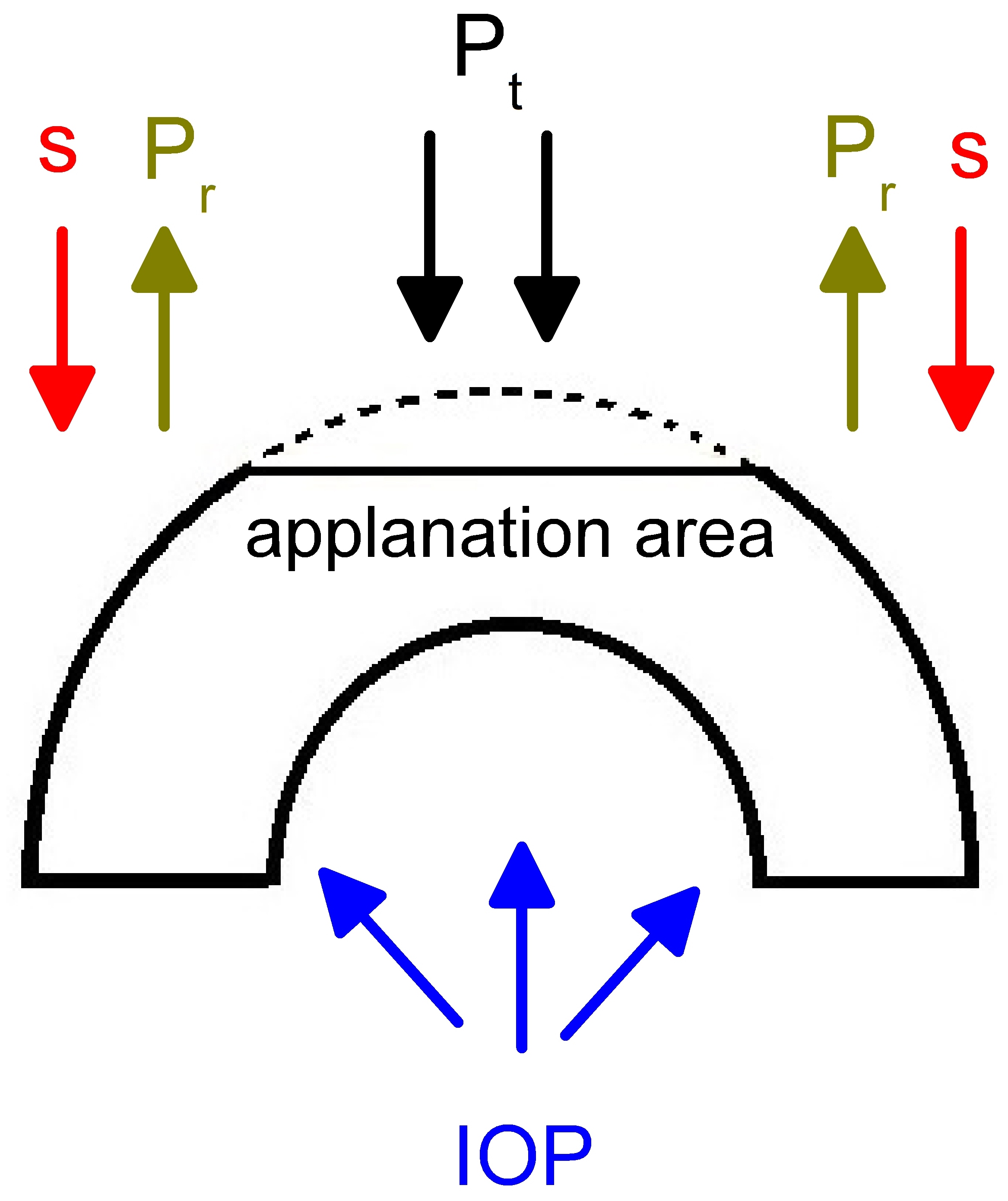}
\caption{Pressures acting during applanation tonometry.
The intraocular pressure IOP on the posterior surface of the cornea
is subtracted to the sum of the tonometer applied pressure
$P_{\rm t}$ and the tear film pressure $s$ so as to obtain the resultant
intraocular pressure at applanation $P_{\rm r}$.}
\label{fig2}
\end{center}
\end{figure}

Let us now analyze how the different
pressures act on the anterior and
posterior corneal surfaces at
applanation (please, see figure \ref{fig2}).
The intraocular pressure IOP on
the posterior surface of the cornea
is subtracted to the sum of the
tonometer applied pressure
$P_{\rm t}(t_{\rm ap,i})$
and the tear film pressure $s$,
so as to obtain the resultant
intraocular pressure at applanation
$P_{\rm r}(t_{\rm ap,i})$
(Liu and Roberts 2005,
Glass \emph{et al.} 2008,
Kotecha \emph{et al.} 2015)
\begin{equation}\label{resultantIOP}
P_{\rm r}(t_{\rm ap,i}) =
P_{\rm t}(t_{\rm ap,i})+s-\textrm{IOP},
\quad \textrm{for} \ i=1,2.
\end{equation}
This radial stress $P_{\rm r}(t_{\rm ap,i})$
can be related to the membrane stress
$\sigma(t_{\rm ap,i})$ in the KVM
via the Laplace law (Glass \emph{et al.} 2008)
\begin{equation}\label{memstress}
\sigma(t_{\rm ap,i})= \frac{R_{\rm c}}{2e}
P_{\rm r}(t_{\rm ap,i}),
\quad \textrm{for} \ i=1,2,
\end{equation}
where $R_{\rm c}$ and $e$ state for the corneal
radius of curvature and corneal
thickness, respectively.
Therefore, introducing equation (\ref{memstress})
into (\ref{taucor}) and performing some
elementary calculations, we derive the final
expression for the corneal retardation time
\begin{equation}\label{taucorp}
\tau = \frac{2\left(|P_1| - |P_2|\right)}
{|\dot{P}_1|+|\dot{P}_2|} = \frac{2 \
\textrm{CH}}{|\dot{P}_1|+|\dot{P}_2|},
\end{equation}
where $P_i=P_{\rm t}(t_{\rm ap,i})$ are the
tonometer applanation pressures,
$\textrm{CH}=|P_1| - |P_2|$ the corneal hysteresis
and $\dot{P}_i$ the tonometer applanation pressures rates.
One observes that the corneal retardation
time $\tau$ is directly related to $\textrm{CH}$,
but with a clear different behaviour and physical
meaning, as will be examined in the next sections.

\section{Methods}

To our purpose, a total number of 200 eyes
from 100 healthy European Caucasian subjects
(mean age $24 \pm 5$ years old) were involved
in the study. Within this group, about 10\% had
an elevated IOP with more than 21 mmHg,
and only patient number $\#278$ had been
undergoing medical treatment for elevated
IOP and diagnosed glaucoma disease
during the measurements. The inclusion
criterion was to be aged between 18
to 30 years old, whereas subjects with
history of ocular pathologies, corneal injuries
or surgery, contact lens wearers
or irregular astigmatism were excluded.
This study was reviewed by an ethical review
board and conforms to the tenets of the Declaration
of Helsinki (Ethical Committee of Research
of the Health Sciences Institute of
Arag\'{o}n, Spain) approved with reference
C.P.-C.I.PI20/377. All participants were
informed about the nature of the project
and signed an informed consent document.

Hence, participants involved
in this study were divided
into two groups: control (with intraocular
pressure values less than 21 mmHg) and ocular
hypertensive (where $\textrm{IOP}_{\rm cc}$
is greater or equal than 21 mmHg),
all of them (as previously mentioned)
healthy subjects without ophthalmological
clinical manifestations, except for patient number
$\#278$ with diagnosed glaucoma disease
(please, see Table \ref{table1}).
This differentiation will be more necessary
and evident when we study the corneal
retardation time as a function of the intraocular pressure,
as described in figure \ref{fig7} of the next section.
\begin{table}[ht!]
\centering
\begin{tabular}{|c|c|}
\hline
{\normalsize control population}
({\footnotesize $\textrm{IOP}_{\rm cc}<21$ mmHg}) &
{\normalsize ocular hypertensive}
({\footnotesize $\textrm{IOP}_{\rm cc}
\ge 21$ mmHg}) \\ [0.5ex]
\hline\hline
90 subjects & 10 subjects \\ [1ex]
$\textrm{mean} \ \textrm{IOP}_{\rm cc} =
(16.38 \pm 2.52)$ mmHg &
$\textrm{mean} \ \textrm{IOP}_{\rm cc} =
(22.43 \pm 1.10)$ mmHg \\ [1ex]
\hline
\end{tabular}
\caption{Control population and OHT subjects
that participated in our study.}
\label{table1}
\end{table}

The applanation pressure data were collected
with the non-contact tonometer Ocular Response
Analyzer ($\textrm{ORA}^{\textsuperscript
{\textregistered}}$; Reichert Ophthalmic
Instruments, Depew, NY) which measures,
apart from the Goldmann-correlated IOP
($\textrm{IOP}_{\rm g}$) and the
corneal-compensated IOP ($\textrm{IOP}_{\rm cc}$),
some biomechanical properties
such as the corneal hysteresis ($\textrm{CH}$)
(related to the capacity
of the cornea to absorb and dissipate
energy) or the corneal resistance factor
($\textrm{CRF}$). This last metric is thought
to be a better indicator of the corneal
viscoelasticity than CH (Gatinel 2007).

The ORA device generates a 25 ms collimated air jet
to deform the cornea and uses an infrared
(IR) detection system in which the IR
emitter is aligned on one side of the
cornea with an IR detector (Roberts 2014).
As the cornea deforms under the applied
air pressure, it rapidly traverses a
state of applanation, causing the
reflected IR light to align with the
detector. As a result, the captured light
increases significatively and a spike
in the IR signal is recorded. Hereafter,
the cornea takes on a slight concave shape,
to then move outward in another applanation
state. Finally, the cornea recovers its
normal configuration state.
In our study, four measures were carried out for
each subject's eye in order to get averaged values.

\begin{figure}
\begin{center}
\includegraphics[width=.65\linewidth]{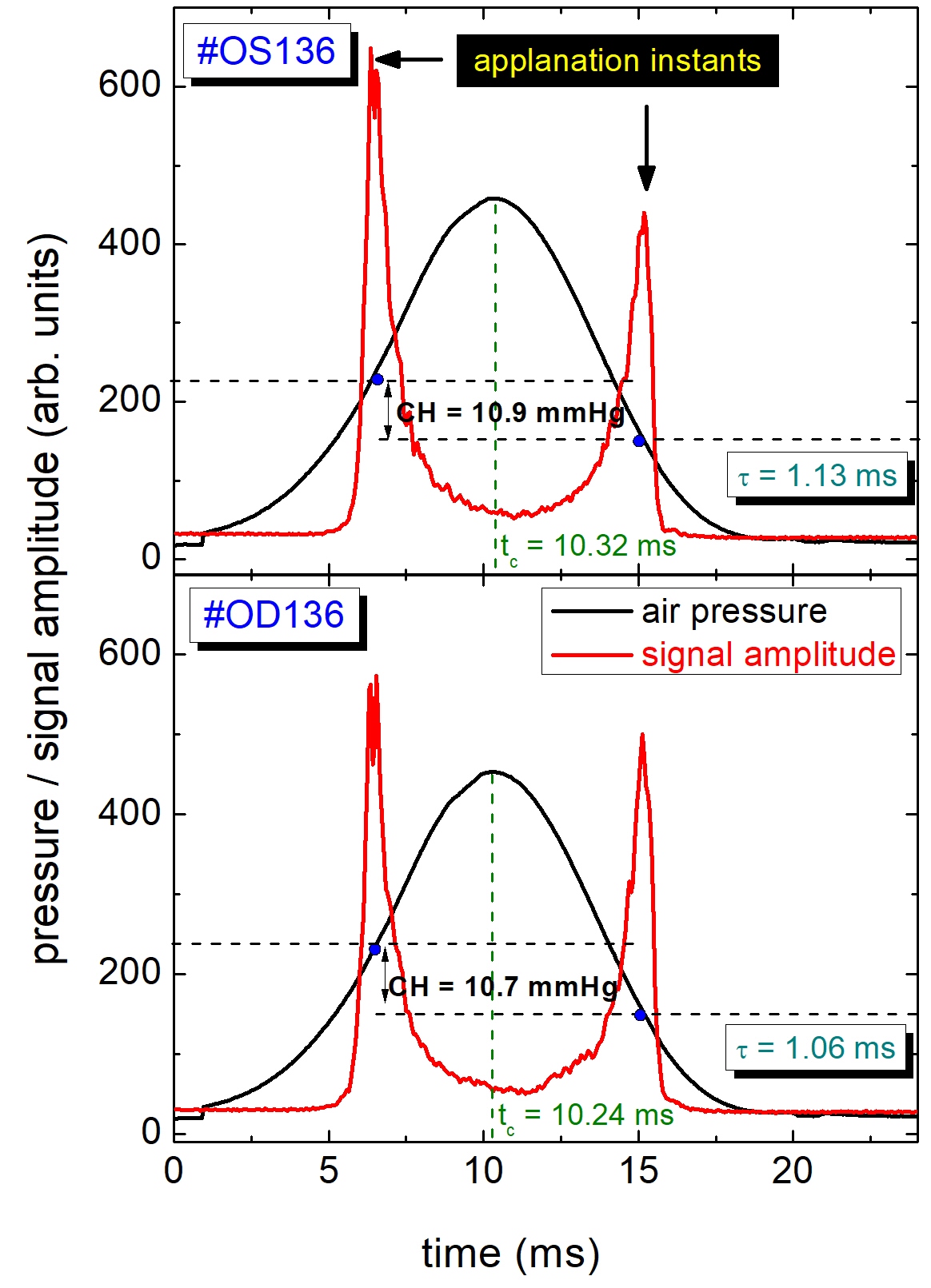}
\caption{ORA's pressure and signal amplitudes for subject
$\#136$ in our population, where the upper (bottom)
panel shows the results for the left (right) eye.
The gaussian center $t_{\rm c}$ (located near
the corneal concave state), corneal hysteresis
$\textrm{CH}$ and corneal retardation time
$\tau$ (as calculated via
equation (\ref{taucorp})) are also shown.}
\label{fig3}
\end{center}
\end{figure}

In this sense, the accurate method to
determine the corneal retardation time $\tau$
was performed via the two ORA's characteristic
curves: the signal amplitude (corresponding
to the IR light which is reflected off
the surface of the cornea during perturbation)
and the pressure amplitude (i.e., the external
applied pressure $P_{\rm t}(t)$
as depicted in figure \ref{fig2}).
The last curve can be fairly fitted
by the following gaussian profile
($0.985 < R^{2} < 0.997$ in all cases)
\begin{equation}\label{Ptgauss}
P_{\rm t}(t) = P_{\rm t,0} + \exp\left[-0.5
\left(\frac{t-t_{\rm c}}{\Delta t}\right)^{2}\right],
\end{equation}
where $\Delta t$ stands for the
pressure amplitude width, and the gaussian center
$t_{\rm c}$ is located near the corneal
concave state. Hence, we have represented
in figure \ref{fig3} both the pressure and
signal amplitudes versus time for subject $\#136$
in our population, where the upper (bottom)
panel shows the results for the left (right) eye.
As easily noticed, the pressure curve
conforms a clear gaussian shape in accordance
with equation (\ref{Ptgauss}), where the gaussian center
$t_{\rm c}$ is also depicted.

Once this gaussian fit is performed,
the signal amplitude provides the two
applanation pressures $P_1$ and $P_2$
(via the sharp peaks in both panels)
and the corresponding corneal hysteresis
$\textrm{CH}$ (which resulted to be 10.9 mmHg
for the left eye and 10.7 mmHg for the right eye,
respectively). The pressure rates $\dot{P}_i$ can be
analytically evaluated from the first derivatives
of the tonometer pressures. Ergo,
the corneal retardation time $\tau$ is
calculated via equation (\ref{taucorp})
(for subject $\#136$, this parameter was
1.13 (1.06) ms for the left (right) eye,
respectively). These corneal retardation time results
are fairly close to the average mean of
the total population, as explained
in detail in the next section.

\section{Results}

In this section we deal with the fundamental
results concerning the corneal retardation
time of our young population, and its important
relationship with the intraocular pressure IOP.

\begin{figure}
\begin{center}
\includegraphics[width=.75\linewidth]{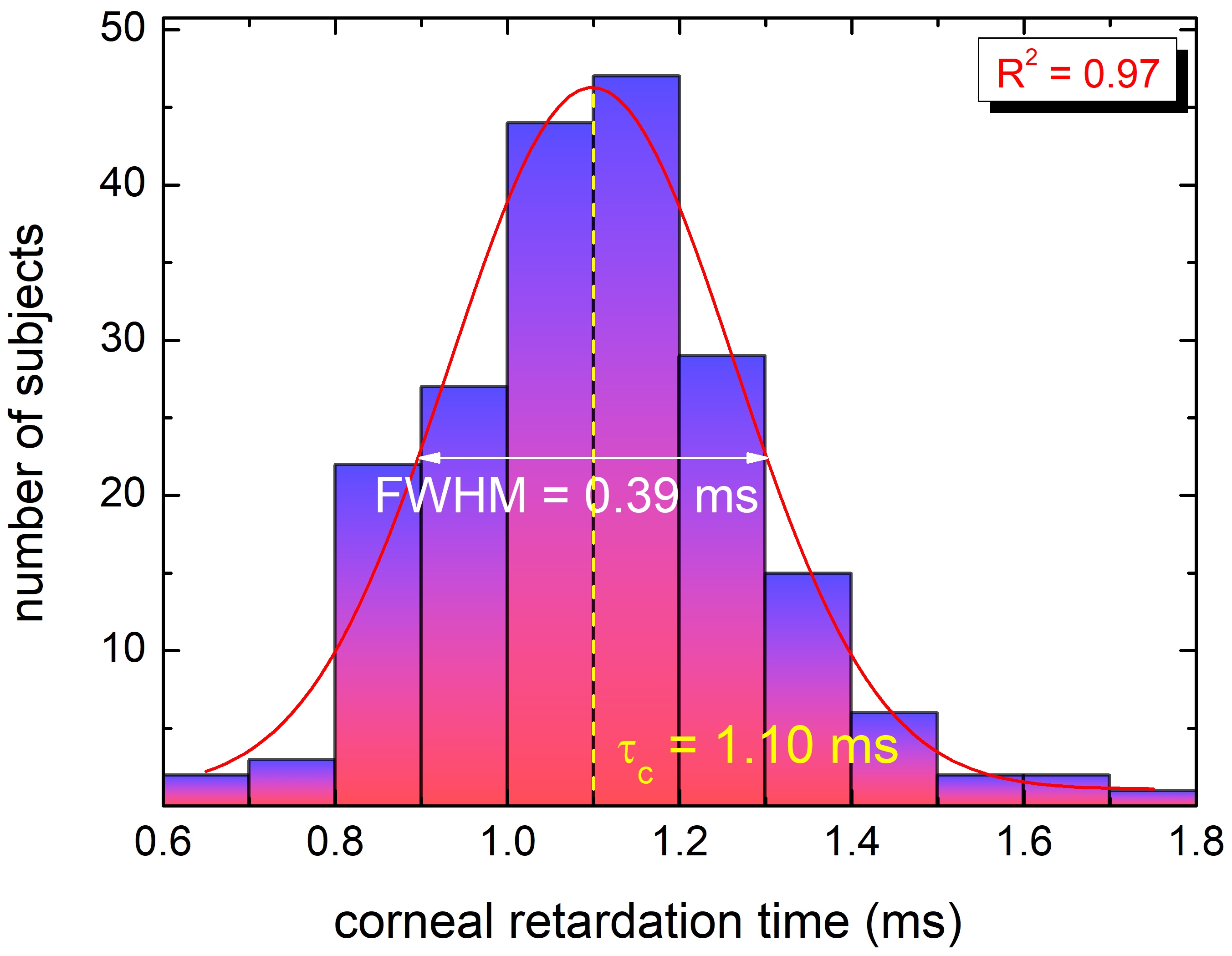}
\caption{Histogram representation of
the corneal retardation time $\tau$
of our study population (200 eyes from
100 young subjects) calculated
via equation (\ref{taucorp}). A clear gaussian
shape ($R^{2}=0.97$) centered at 1.10 ms
and FWHM of 0.39 ms is found. Elastic
corneas are associated with low $\tau$ values
(about 13.5\% in our case) while
elevated corneal retardation times are related
to more viscoelastic corneas (roughly a 13\% of
our population).}
\label{fig4}
\end{center}
\end{figure}

To this aim, the histogram illustrated
in figure \ref{fig4} shows the corneal
retardation time $\tau$ (calculated via
equation (\ref{taucorp})) of the 100 subjects
that participated in the study. It can be observed
an explicit gaussian profile centered at 1.10 ms
with a full width half maximum (FWHM) of 0.39 ms.
The R-squared parameter for this gaussian fit was 0.97.
In view of these results it can be assumed that
for a young and healthy population, the corneal
retardation time should be ranged
between 0.90 and 1.30 ms, where more elastic
corneas are associated with low $\tau$ values
(about 13.5\% in our case). In addition,
elevated corneal retardation times are related
to viscoelastic corneas (roughly a 13\% of
our population), not necessarily
being pathological cases those subjects with
upper or lower $\tau$ values (though, in
the later scenario, a clear connection with
higher intraocular pressures is found,
as briefly discussed).

\begin{figure}
\begin{center}
\includegraphics[width=.75\linewidth]{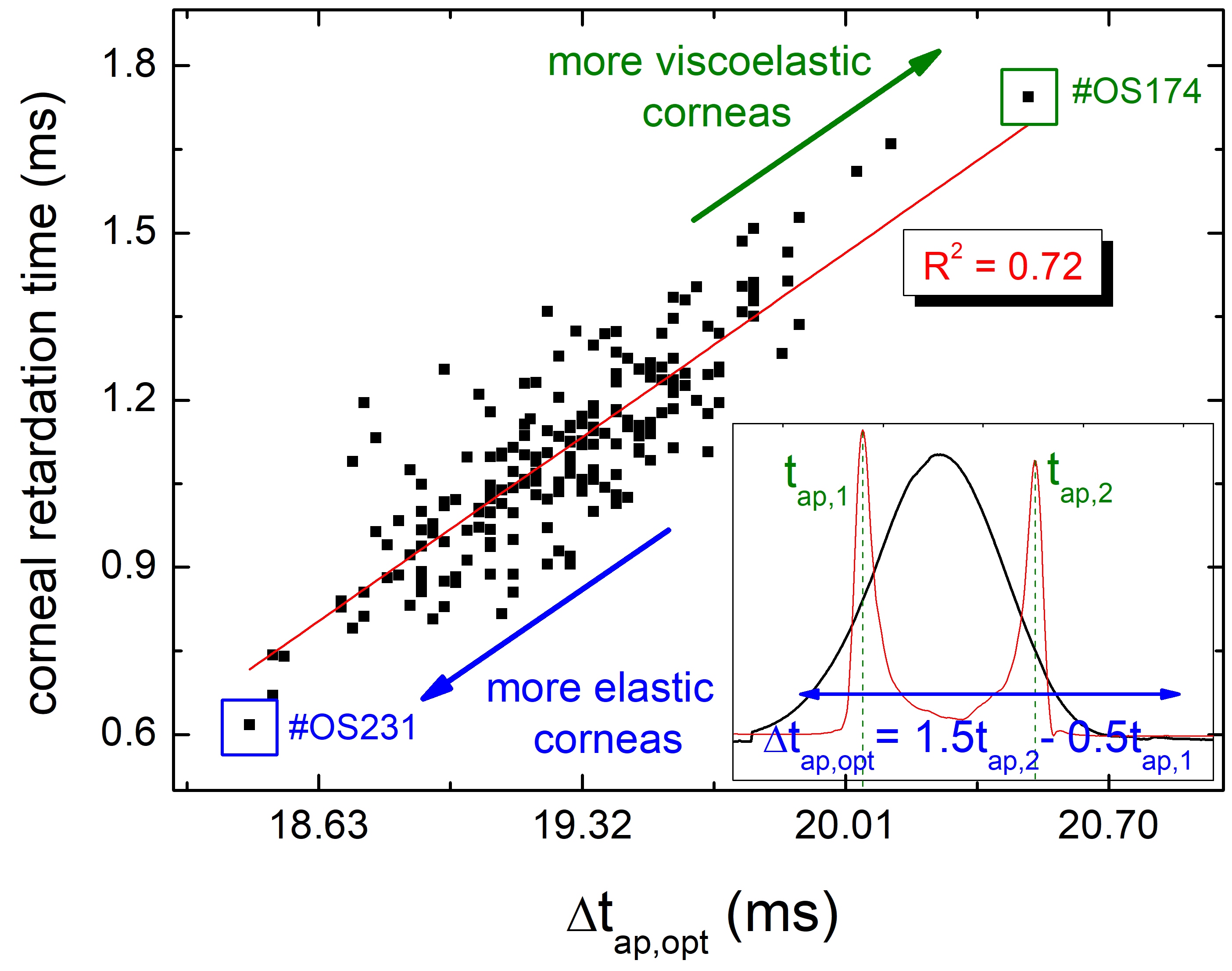}
\caption{Corneal retardation time $\tau$
of our study population versus the optimized
time interval $\Delta t_{\rm ap,opt}$
(over which the biomedical data are highly correlated,
with $R^{2}=0.72$). The inset shows more clearly
this optimized applanation lapse. It can also
be appreciated that lower (higher) $\tau$ values
correspond to more elastic (viscoelastic)
corneas.}
\label{fig5}
\end{center}
\end{figure}

On the other hand, it is expected that
our biomechanical parameter should be directly
correlated to the time interval between the
two applanation times $\Delta t_{\rm ap}$.
That is, more elastic (viscoelastic) corneas,
which entail lower (higher) $\tau$ values,
might take less (more) time during
the applanation interval. In such a case
(not shown in this work), the linear coefficient
of determination resulted to be $R^{2}=0.46$
for $\Delta t_{\rm ap} = t_{\rm ap,2} - t_{\rm ap,1}$.
Subsequent data analysis (please,
see figure \ref{fig5}) demonstrated that the
optimized time interval corresponded to
$\Delta t_{\rm ap,opt} = 1.5 t_{\rm ap,2}
- 0.5 t_{\rm ap,1}$, where now $R^{2}=0.72$.
This time lapse is depicted in the
inset of figure \ref{fig5}, however different
applanation time intervals might also be
considered for our study (with $R^{2} > 0.45$,
in all cases). As a matter of fact,
the optimized time interval for
subject $\#\textrm{OS}231$
was 18.46 ms, fairly shorter than
patient $\#\textrm{OS}174$ with a
time lapse of 20.49 ms. This may be
interpreted assuming that the
cornea of the former subject is
more elastic (that is, it takes
less time between both applanation
times) than subject $\#\textrm{OS}174$,
with a more viscoelastic cornea.
Moreover, these findings should
be affected by the intraocular pressure,
because elastic corneas with low IOP values
might take longer to recover its original shape
than viscoelastic corneas of OHT subjects.
This relationship between the $\tau$ parameter
and the IOP will be treated in detail shortly.

\begin{figure}
\begin{center}
\includegraphics[width=.75\linewidth]{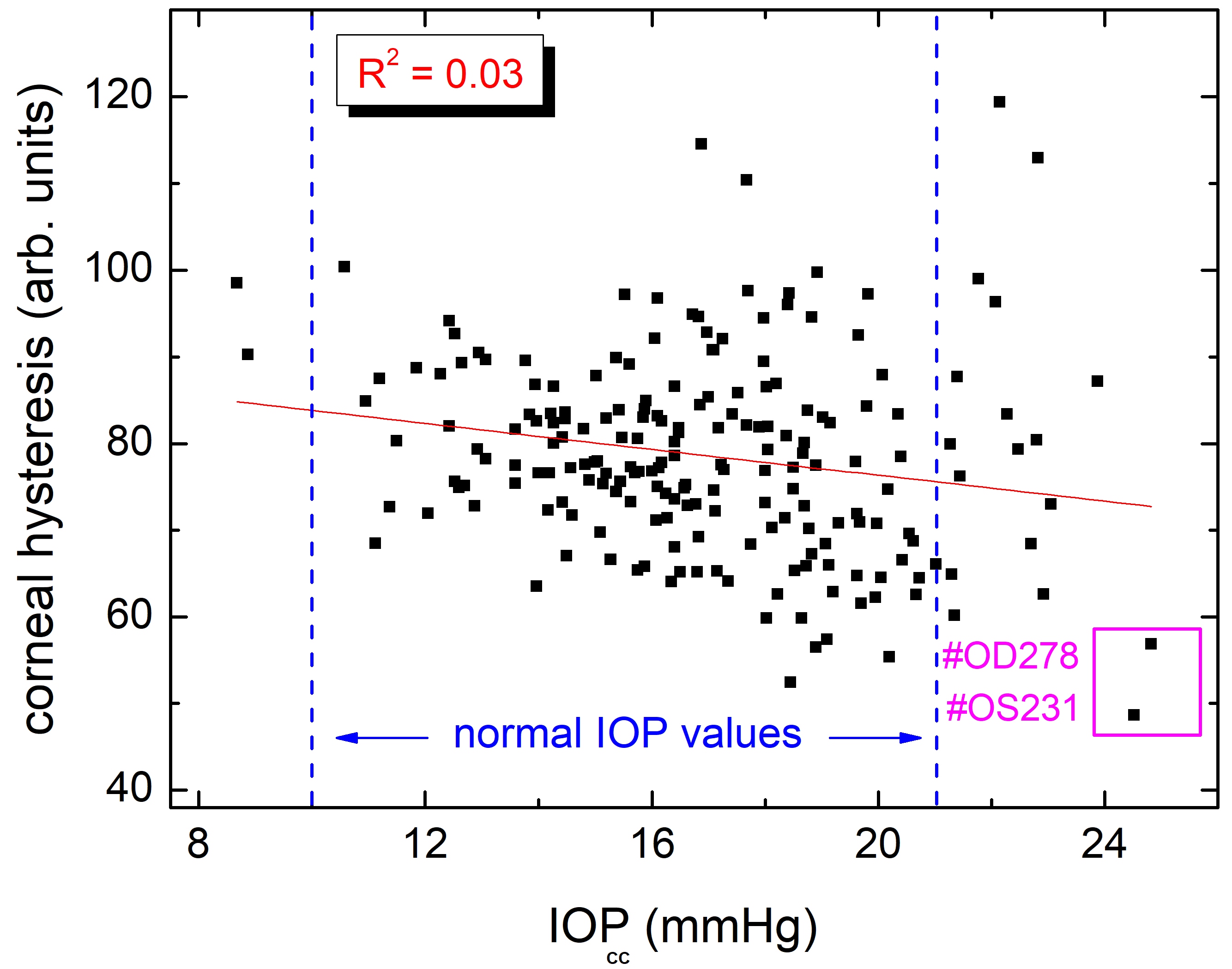}
\caption{Corneal hysteresis $\textrm{CH}$
as a function of the corneal-compensated
intraocular pressure $\textrm{IOP}_{\rm cc}$
for our study population. It can be noticed
that both parameters are highly uncorrelated,
in consistency with previous reported
research (Luce 2005). On the other hand,
low $\textrm{CH}$ subjects like
$\#\textrm{OD}278$ or $\#\textrm{OS}231$
possess high intraocular pressures and an
possible risk of glaucoma progression.}
\label{fig6}
\end{center}
\end{figure}

But before embarking on this study,
let us first analyze an important biomechanical
parameter like the corneal hysteresis
$\textrm{CH}$ and its dependence on the
corneal-compensated $\textrm{IOP}_{\rm cc}$
in our population (please, see figure \ref{fig6}).
Assuming that normal IOP ranges from 10 to 21 mmHg
(Badakere \emph{et al.} 2021),
one notices that both parameters
are not correlated, in consistency with previous
published work (Luce 2005), where no statistical
significance was found. For this reason, we
have not differentiated between control and
OHT populations. Nevertheless, it can be observed
that low $\textrm{CH}$ values (such as subjects
$\#\textrm{OD}278$ or $\#\textrm{OS}231$) also
possess high intraocular
pressures and an possible risk of glaucoma progression.
This result agrees with prior reported research,
where low corneal hysteresis is thought to be related to
the risk and development of glaucoma
(Prata \emph{et al.} 2012, Deol \emph{et al.} 2015),
though there is no consensus on this topic.
As stated by Roberts (2014), low $\textrm{CH}$ should
not be interpreted as a damaged cornea,
and further work is required to determine what
component contributing to this viscoelastic
parameter correlates to damage at the optic nerve.
In our study, when the corneal hysteresis
$\textrm{CH}$ is divided by the sum of the
first derivatives of the applanation pressures
$\dot{P}_i$ (please, see again equation (\ref{taucorp})),
the uncorrelated scheme illustrated in
figure \ref{fig6} turns to a well-defined linear
dependence, as immediately discussed.

\begin{figure}
\begin{center}
\includegraphics[width=.90\linewidth]{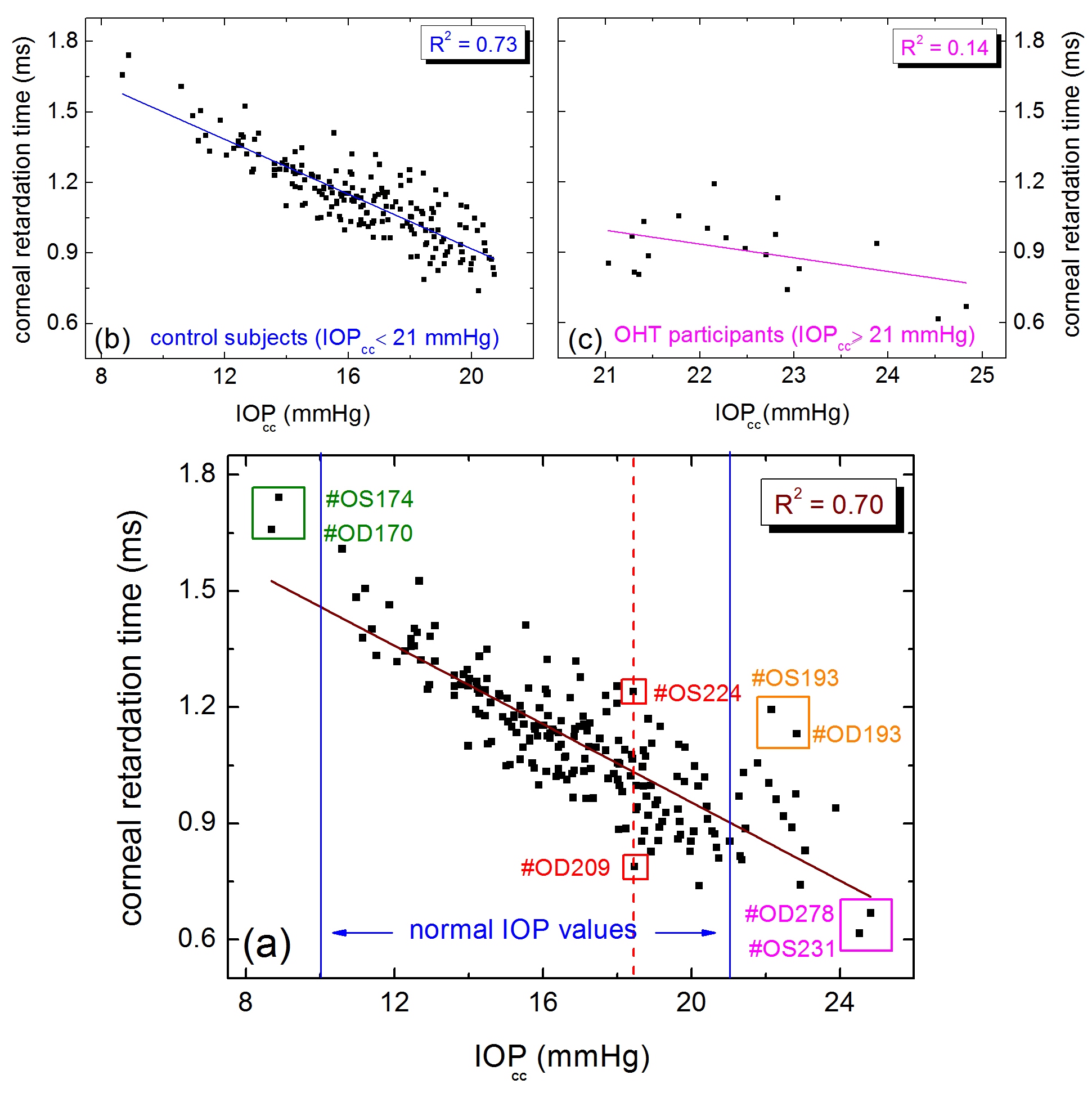}
\caption{(a) Corneal retardation time $\tau$
of our study population (calculated via
equation (\ref{taucorp})) as a function of the
corneal-compensated $\textrm{IOP}_{\rm cc}$
given by ORA. It can be observed that
both parameters are highly correlated with
a linear coefficient of determination $R^{2}=0.70$.
On the other hand, it can be argued that
OHT subjects like $\#\textrm{OD}278$ (who was receiving
medication for elevated IOP and possess
a particularly elastic cornea with $\tau=0.67$ ms)
should be more likely to develop glaucoma than
other OHT subjects with more viscoelastic
corneas (like patient $\#\textrm{OS}193$
in our population). Furthermore, the upper
panels represent the $\tau$ parameter for
(b) our control population and (c) OHT participants.
A clear linear correlation is found for the control
group, while no significant correlation is
obtained for our OHT population.}
\label{fig7}
\end{center}
\end{figure}

Accordingly, our fundamental result is
exhibited in figure \ref{fig7}(a) where the
corneal retardation time $\tau$ is represented
as a function of $\textrm{IOP}_{\rm cc}$.
Now, a clear linear dependence is found
(with a coefficient of determination $R^{2}=0.70$),
where lower $\tau$ values are mostly associated with
higher intraocular pressures. This can be easily
understood since more pressurized corneas will
behave more elastically than those with lower IOPs.
So, for instance, subjects $\#\textrm{OD}170$
or $\#\textrm{OS}174$ have the highest
$\tau$ values in our study (which should
be connected with small IOPs) but
it cannot be assured that such corneas
are the most viscoelastic of our population.
This fact should be corroborated with a
relevant number of ocular hypotony patients,
though it does not constitute a subject
of study in our current research.

Nonetheless, this is not a fundamental rule.
Indeed, after inspection of figure \ref{fig7}(a),
one notices that for similar $\textrm{IOP}_{\rm cc}$ values
(such as for subjects $\#231$, $\#193$ or $\#278$),
the corneal viscoelastic behavior is different.
Whereas subject $\#193$ possess good viscoelastic corneas
for both eyes (greater than the average of 1.10 ms,
as illustrated in figure \ref{fig4}), other OHT patients
like $\#231$ or $\#278$ have more elastic corneas
(that is, with lower $\tau$ values). This means that
the corneas of subject $\#193$ would be more prepared
to absorb IOP fluctuations (and avoid possible
glaucoma progression) than the other OHT patients.
As a consequence, hypertensive subjects with low
corneal retardation times should be periodically monitored,
in order to prevent possible optical nerve damage.
Moreover, even normal IOP subjects present significant
differences between their $\tau$ values, although
these patients are not considered as a
''risky population''. In this sense, at a normal
IOP of 18.4 mmHg (please note the vertical dashed
line in figure \ref{fig7}), subject $\#\textrm{OD}209$
exhibits a more elastic cornea (0.79 ms)
than patient $\#\textrm{OS}224$, where the
corneal retardation time resulted to be 1.24 ms
(a 57\% higher than the latter). Additionally,
the corneal elasticity of subject $\#\textrm{OD}209$
(quantified by our $\tau$ parameter)
is similar to some OHT patients in our study,
so such normal IOP subjects should also be
controlled, despite they do not
belong to a risk group.

For the sake of clarity,
we have also illustrated the corneal retardation
time $\tau$ for the control population
(figure \ref{fig7}(b)) and OHT subjects
(figure \ref{fig7}(c)). Clearly,
the linear correlation between the $\tau$
parameter and the intraocular pressure
$\textrm{IOP}_{\rm cc}$ for the control group
is even increased (as compared to the whole
population), while no significant correlation
for OHT subjects is found. This fact reflects the
difficulty in predicting the corneal
retardation time for our ocular hypertensive
population, probably due to the small number
of OHT participants in our study (please, see again
Table \ref{table1}).

\section{Discussion}

Summarizing, a detailed analysis of the corneal
retardation time $\tau$ of a young population
(200 eyes from 100 healthy subjects) has been
carried out. Our results show that this parameter
is highly correlated with the corneal-compensated
intraocular pressure $\textrm{IOP}_{\rm cc}$
supplied by ORA tonometer, underlying the risk
for OHT subjects with lower $\tau$ values to
develop hypertension diseases (due to the
inability of the poorly viscoelastic cornea
to absorb IOP fluctuations). Indeed, viscous
damping of the cornea should be crucial since
increased damping capacity of the eye may
actually buffer hazardous IOP fluctuations,
diminishing the stress/strain on the
optic nerve and peripapillary scleral tissues
(Kaushik and Pandav 2012).

Furthermore, some authors argue that
$\textrm{IOP}_{\rm cc}$ is overestimated
(Martinez \emph{et al.} 2006) in comparison
with the gold standard technique in measuring IOP,
that is, the Goldmann applanation tonometry
(Lee \emph{et al.} 2018). Thus, a possible
discrepancy between our results for
the $\tau$ parameter and the
corneal-compensated intraocular pressure
should be expected. However, given that all
$\textrm{IOP}_{\rm cc}$ values in our population
might be affected by the same (or similar) scale factor,
the linear dependence depicted in figure \ref{fig7}
should remain the same, with comparable
R-squared parameters.

As previously stated, the fundamental aim
of our work is to yield an useful tool (i.e.,
the corneal retardation time $\tau$) to
systematically discern which ocular hypertensive
patients are more likely to develop OHT diseases
and ensure an early diagnosis. Among them,
glaucoma plays a leading role since this eye illness
is the most common cause of irreversible blindness
and affects about 80 million people worldwide,
with many more undiagnosed (Tribble \emph{et al.} 2023).

In this sense, our work might help in glaucoma diagnosis
(as compared to previous existing methods already mentioned
in the Introduction) due to the easiness and robustness of our
method. More specifically, it is straightforward to measure the
corneal retardation time $\tau$ with non-contact tonometers
(i.e., via our equation (\ref{taucorp})
and the applanation pressures provided by such instruments).
Besides, these values are not affected by subjective
or qualitative factors, so the $\tau$ parameter can be considered
a strong biomechanical indicator not subject
to intra- or interobserver variabilities. Also, it has been
suggested in the literature that increased viscoelasticity
of ocular tissues may have a protective role in
glaucoma (Murphy \emph{et al.} 2017, Del Buey \emph{et al.} 2021),
so the $\tau$ parameter could be an important metric to
diagnose this disease, playing a leading role in
explaining normotensive glaucoma. Instead,
our method does not provide information about
structural changes in the optic nerve head or the retinal nerve
fiber layer, so a proper glaucoma diagnosis might
not be guaranteed (and other techniques such as
the fundus photography or optical coherence tomography
are needed).

It has been widely reported in the literature that
the major risk factors for glaucoma are genetics, age
and an elevated IOP, so, intraocular pressure should be
adequately controlled to avoid visual field deterioration.
Beyond the diagnosis, the prediction of the
future glaucoma progression of an individual patient
is often extremely difficult for clinicians, due to
the mix of the abovementioned risk factors.
In this regard, corneal
thickness, corneal hysteresis or horizontal and vertical
cup-disc ratio constitute additional OHT risk factors
(apart from an elevated IOP). As reported by Murphy
\emph{et al.} (2017), about 30–50\% of glaucoma patients
have normal IOP values, becoming evident that
other elements should be taken into account.
Provided the large number of independent risk factors,
it might not seem plausible that a unique parameter
as the corneal retardation time would effectively
serve as a OHT indicator.

Nevertheless, the validity of
our theoretical approach (as given by the fundamental
equation (\ref{taucorp}) for the $\tau$ parameter)
is based on the strength of the Kelvin-Voigt
model to imitate the corneal viscoelastic behaviour.
In other words, the $\tau$ metric constitutes
a valuable indicator of the corneal viscoelastic quality.
However, its validity is subject to obtaining well-defined
signals via non-contact tonometers (please, see again
figure \ref{fig3}): irregular signals with no evident
applanation peaks will not give reliable corneal
retardation time values.

In fact, our proposal is consistent with previous
reported glaucoma research (Matsuura \emph{et al.} 2017)
where it is suggested that careful
consideration should be given to patients
whose eyes are applanated fast in the first and second
applanations (please, see again
our fundamental results concerning the applanation
time interval in figure \ref{fig5}, which are directly
related to the $\tau$ parameter shown in figure \ref{fig7}(a)).
Additionally, the loss of corneal viscoelasticity
(which is correlated with lower $\tau$ values in our
model) is a risk factor that can lead an ocular hypertensive
subject to develop glaucoma disease
(Roberts \emph{et al.} 2023). For that matter,
the corneal retardation time might represent an early detector of
those complications associated with ocular hypertension
before clinical signs manifest.
Nonetheless, such suspected glaucoma patients should
be periodically monitored to confirm this fact.

Among the potential limitations encountered
during our study, we can mention the reduced number
of OHT subjects (in comparison with the normal IOP patients)
and the difficulty in finding some OHT clinical cases.
It is worth mentioning that our young study population consisted of
100 healthy subjects and, within this group,
about 10\% had an elevated IOP with more than 21 mmHg
(only patient $\#278$ was medically monitored
due to its high IOP). Certainly, an increased number of
OHT subjects is needed for a conclusive
statement about the utility of the corneal
retardation time as an ocular hypertension
disease indicator. Moreover, it could be interesting to include
participants with pathological conditions such as diabetic retinopathy
or glaucoma for a more comprehensive analysis, as well as considering
another age group. In addition, it should be studied the effects
of physical parameters such as central corneal thickness (CCT)
or corneal morphology on corneal biomechanics (in line with
the proposals by Marcell\'{a}n \emph{et al.} 2022)
which is a subject of ongoing research by our group.

\ack{Oscar del Barco gratefully thanks
Alfonso Jimenez Villar for helpful discussions on
1D corneal rheological models. The authors
acknowledge the funding grant from
Departamento de Ciencia, Universidad
y Sociedad del Conocimiento del Gobierno de
Arag\'{o}n (research group $\textrm{E44-23R}$).}

\section*{Data availability statement}

The datasets generated and analyzed during the current
study are not publicly available because it contains
biometric data of participants and cannot be published,
but are available from the corresponding author
on reasonable request.

\section*{Conflicts of interest}

The authors declare no conflict of interest.

\section*{ORCID iDs}

Oscar del Barco https://orcid.org/0000-0001-7502-9164 \\
Francisco J \'{A}vila https://orcid.org/0000-0002-9068-7728 \\
Concepci\'{o}n Marcell\'{a}n https://orcid.org/0000-0002-7516-3029 \\
Laura Rem\'{o}n https://orcid.org/0000-0002-3979-4528 \\

\bigbreak
\section*{References}

\begin{harvard}

\item[] Ambr\'{o}sio R \emph{et al.} 2017
Integration of scheimpflug-based corneal tomography and
biomechanical assessments for enhancing ectasia detection
{\it J. Refract. Surg.} \textbf{33(7)} 434-443

\item[] Anderson D R 2003
Collaborative normal tension glaucoma study
{\it Curr. Opin. Ophthalmol.} \textbf{14(2)} 86-90

\item[] Ariza-Gracia M A \emph{et al.} 2015
Coupled biomechanical response of the cornea assessed by
non-contact tonometry. A simulation study
{\it PLoS ONE} \textbf{10(3)} e0121486

\item[] Asejczk-Widlicka M \emph{et al.} 2019
Data analysis of the ocular response analyzer for
improved distinction and detection of glaucoma
{\it J. Opt. Soc. Am. A} \textbf{36(4)} B71-B76

\item[] Badakere S V \emph{et al.} 2021
Agreement of Intraocular Pressure Measurement
of Icare ic200 with Goldmann Applanation Tonometer
in Adult Eyes with Normal Cornea
{\it Ophthalmol. Glaucoma} \textbf{4} 238-243

\item[] Brinson H F and Brinson L C 2008
Polymer Engineering Science and Viscoelasticity.
Springer, Berlin

\item[] Brown L \emph{et al.} 2018
The Influence of Corneal Biomechanical
Properties on Intraocular Pressure Measurements
Using a Rebound Self-tonometer
{\it J. Glaucoma} \textbf{27(6)} 511-518

\item[] Catania F \emph{et al.} 2023
Corneal Biomechanics Assessment with Ultra
High Speed Scheimpflug Camera in Primary
Open Angle Glaucoma Compared with Healthy
Subjects: A meta-analysis of the Literature
{\it Curr. Eye Res.} \textbf{48:2} 161-171

\item[] Chakrabarti L \emph{et al.} 2016
Automated Detection of Glaucoma from Topographic
Features of the Optic Nerve Head in Color Fundus Photographs
{\it J. Glaucoma} \textbf{25} 590-597

\item[] Chan E \emph{et al.} 2021
Changes in Corneal Biomechanics and Glaucomatous
Visual Field Loss
{\it J. Glaucoma} \textbf{30(5)} 246-251

\item[] Consejo A \emph{et al.} 2019
Corneal Properties of Keratoconus Based
on Scheimpflug Light Intensity Distribution
{\it Investig. Ophthalmol. Vis. Sci.} \textbf{60} 3197–3203

\item[] De Moraes C G \emph{et al.} 2012
Effect of treatment on the rate of visual
field change in the ocular hypertension treatment
study observation group
{\it Invest. Ophthalmol. Vis. Sci.} \textbf{53(4)} 1704-1709

\item[] Del Buey-Sayas M A \emph{et al.} 2021
Corneal Biomechanical Parameters and Central Corneal
Thickness in Glaucoma Patients, Glaucoma Suspects,
and a Healthy Population
{\it J. Clin. Med.} \textbf{10(12)} 2637

\item[] Deol M \emph{et al.} 2015
Corneal hysteresis and its relevance to glaucoma
{\it Curr. Opin. Ophthalmol.} \textbf{26(2)} 96-102

\item[] Fraldi M \emph{et al.} 2016
Visco-elastic and thermal-induced damaging in
time-dependent reshaping of human cornea after
conductive keratoplasty
{\it Mech. Time-Depend. Mater.} \textbf{21(1)} 45–59

\item[] Garcia-Porta N \emph{et al.} 2014
Corneal biomechanical properties in different
ocular conditions and new measurement techniques
{\it ISRN Ophthalmol.} \textbf{2014} 724546

\item[] Gatinel D 2007
Evaluating biomechanic properties of the cornea
{\it Cataract. Refract. Surg. Today Eur.} \textbf{25} 36-39

\item[] Geevarghese A \emph{et al.} 2021
Optical Coherence Tomography and Glaucoma
{\it Annu. Rev. Vis. Sci.} \textbf{15;7} 693-726

\item[] Glass D H \emph{et al.} 2008
A Viscoelastic Biomechanical Model of the
Cornea Describing the Effect of Viscosity
and Elasticity on Hysteresis
{\it Invest. Ophthalmol. Vis. Sci.} \textbf{49} 3919-3926

\item[] Gonz\'{a}lez-M\'{e}ijome J M \emph{et al.} 2008
Pilot study on the influence of corneal
biomechanical properties over the short term
in response to corneal refractive therapy for myopia
{\it Cornea} \textbf{27} 421-426

\item[] Grise-Dulac A \emph{et al.} 2012
Assessment of corneal biomechanical properties
in normal tension glaucoma and comparison with
open-angle glaucoma, ocular hypertension, and normal eyes
{\it J. Glaucoma} \textbf{21(7)} 486-489

\item[] Hager A \emph{et al.} 2008
Effect of central corneal thickness and
corneal hysteresis on tonometry as measured by
dynamic contour tonometry, ocular response
analyzer, and Goldmann tonometry in glaucomatous eyes
{\it J. Glaucoma} \textbf{17(5)} 361-365

\item[] Han Z \emph{et al.} 2014
Air puff induced corneal vibrations: theoretical
simulations and clinical observations
{\it J. Refract. Surg.} \textbf{30(3)} 208–213

\item[] Jannesari M \emph{et al.} 2018
Numerical and clinical investigation on the material model
of the cornea in Corvis tonometry tests:
differentiation between hyperelasticity and viscoelasticity
{\it Mech. Time Depend. Mater.} \textbf{23} 373–384

\item[] Jimenez-Villar A \emph{et al.} 2022
Rheological Eye Model to Determine Elastic
and Viscoelastic Properties of the Cornea
and Crystalline Lens
{\it Invest. Ophthalmol. Vis. Sci.} \textbf{63} 2395-A0198

\item[] Kass M A \emph{et al.} 2002
The Ocular Hypertension Treatment Study:
A Randomized Trial Determines That Topical Ocular
Hypotensive Medication Delays or Prevents the
Onset of Primary Open-Angle Glaucoma
{\it Arch. Ophthalmol.} \textbf{120(6)} 701-713

\item[] Kaushik S and Pandav S S 2012
Ocular Response Analyzer
{\it J. Curr. Glaucoma Pract.} \textbf{6(1)} 17-19

\item[] Kelly P 2013
Solid mechanics part I: An introduction to solid mechanics
{\it Solid mechanics lecture notes}
University of Auckland

\item[] Kok S \emph{et al.} 2014
Calibrating corneal material model parameters
using only inflation data: an ill-posed problem
{\it Int. J. Numer. Methods Biomed. Eng.}
\textbf{30(12)} 1460–1475

\item[] Kotecha A \emph{et al.} 2015
Tonometry and Intraocular Pressure Fluctuation
{\it Glaucoma} \textbf{1} 98-108

\item[] Lee S Y \emph{et al.} 2018
Utility of Goldmann applanation tonometry
for monitoring intraocular pressure in glaucoma
patients with a history of laser refractory surgery
{\it PLoS ONE} \textbf{13(10)} e0206564

\item[] Lee K M \emph{et al.} 2019
Association of Corneal Hysteresis
With Lamina Cribrosa Curvature in Primary
Open Angle Glaucoma
{\it Invest. Ophthalmol. Vis. Sci.}
\textbf{60(13)} 4171-4177

\item[] Lemij H G and Reus N J 2008
New developments in scanning laser polarimetry for glaucoma
{\it Curr. Opin. Opthalmol.} \textbf{19} 136–140

\item[] Liu J and Roberts C J 2005
Influence of corneal biomechanical
properties on intraocular
pressure measurement: quantitative analysis
{\it J. Cataract Refract. Surg.} \textbf{31(1)} 146-155

\item[] Liu T \emph{et al.} 2020
Characterization of Hyperelastic Mechanical
Properties for Youth Corneal Anterior central
Stroma Based on Collagen Fibril Crimping Constitutive Model
{\it J. Mech. Behav. Biomed. Mater.} \textbf{103} 103575

\item[] Luce D A 2005
Determining in vivo biomechanical properties
of the cornea with an ocular response analyzer
{\it J. Cararact. Refract. Surg.} \textbf{31} 156-162

\item[] Maczynska E \emph{et al.} 2019
Assessment of the influence of viscoelasticity
of cornea in animal ex vivo model using air-puff
optical coherence tomography and corneal hysteresis
{\it J. Biophotonics} \textbf{12(2)} e201800154

\item[] Marcell\'{a}n M C \emph{et al.} 2022
Corneal hysteresis and intraocular pressure are
altered in silicone-hydrogel soft contact lenses wearers
{\it Int. Ophthalmol.} \textbf{42(9)} 2801-2809

\item[] Martinez J M \emph{et al.} 2006
Ocular Response Analyzer versus Goldmann
Applanation Tonometry for Intraocular Pressure Measurements
{\it Invest. Ophthalmol. Vis. Sci.} \textbf{47} 4410-4414

\item[] Matlach J \emph{et al.} 2019
Investigation of intraocular pressure fluctuation as a risk
factor of glaucoma progression
{\it Clin. Ophthalmol.} \textbf{13} 9-16

\item[] Matsuura M \emph{et al.} 2017
Using CorvisST tonometry to assess glaucoma
progression
{\it PLoS ONE} \textbf{12(5)} e0176380

\item[] Medeiros F A and Weinreb R N 2006
Evaluation of the influence of corneal biomechanical
properties on intraocular pressure measurements using
the ocular response analyzer
{\it J. Glaucoma} \textbf{15(5)} 364-370

\item[] Murphy M L \emph{et al.} 2017
Corneal hysteresis in patients with glaucoma-like optic discs,
ocular hypertension and glaucoma
{\it BMC Ophthalmol.} \textbf{17(1)} 1-8

\item[] Ortiz D \emph{et al.} 2007
Corneal biomechanical properties in normal,
postlaser in situ keratomileusis, and keratoconic eyes
{\it J. Cataract Refract. Surg.} \textbf{33} 1371-1375

\item[] Padmanabhan P and Elsheikh A 2023
Keratoconus: A Biomechanical Perspective
{\it Curr. Eye Res.} \textbf{48:2} 121-129

\item[] Prata T S \emph{et al.} 2012
Association between corneal biomechanical
properties and optic nerve head morphology
in newly diagnosed glaucoma patients
{\it Clin. Experiment. Ophthalmol.} \textbf{40(7)} 682-688

\item[] Roberts C J 2014
Concepts and misconceptions in corneal biomechanics
{\it J. Cararact. Refract. Surg.} \textbf{40} 862-869

\item[] Roberts C J \emph{et al.} 2023
Comparison of elastic and viscoelastic biomechanical
metrics in ocular hypertension and normal controls
{\it Invest. Ophthalmol. Vis. Sci.} \textbf{64(8)} 4720

\item[] Roy A S and Dupps Jr. W J 2011
Patient-Specific Computational Modeling of Keratoconus
Progression and Differential
Responses to Collagen Cross-Linking
{\it Investig. Ophthalmol. Vis. Sci.} \textbf{52} 9174–9187

\item[] S\'{a}nchez P \emph{et al.} 2014
Biomechanical and optical behavior of human corneas before
and after photorefractive keratectomy
{\it J. Cataract Refract. Surg.} \textbf{40(6)} 905–917

\item[] Sharma P \emph{et al.} 2008
Diagnostic tools for glaucoma detection and management
{\it Surv. Ophthalmol.} \textbf{53 (SUPPL1)} S17–S32

\item[] Simonini I \emph{et al.} 2016
Theoretical and numerical analysis of the corneal air puff test
{\it J. Mech. Phys. Solids} \textbf{93} 118-134

\item[] Susanna B N \emph{et al.} 2019
Corneal biomechanics and visual field progression in eyes
with seemingly well-controlled intraocular pressure
{\it Ophthalmology} \textbf{126} 1640-1646

\item[] Torres J \emph{et al.} 2022
Torsional wave elastography to assess the
mechanical properties of the cornea
{\it Sci. Rep.} \textbf{12} 8354

\item[] Tribble J R \emph{et al.} 2023
Neuroprotection in glaucoma: Mechanisms beyond intraocular
pressure lowering
{\it Mol. Aspects Med.} \textbf{92} 101193

\item[] Whitford C \emph{et al.} 2018
A Viscoelastic Anisotropic Hyperelastic Constitutive
Model of the Human Cornea
{\it Biomech. Model. Mechanobiol.} \textbf{17} 19–29

\item[] Yaghoubi M \emph{et al.} 2015
Confocal scan laser ophthalmoscope for diagnosing glaucoma:
A systematic review and meta-analysis
{\it Asia Pac. J. Ophthalmol.} \textbf{4} 32–39

\end{harvard}

\end{document}